\documentclass[aps,prd,tighten,11pt,showpacs,a4,superscriptaddress,nofootinbib]{revtex4}

\usepackage{graphicx}
\usepackage{bm}
\usepackage{amsmath,amssymb}
\usepackage{latexsym}

\newcommand{\rp}{r_{+}}

\newcommand{\pp}{\phantom}

\newcommand{\al}{\alpha}

\begin{document}

\title{Inside the degenerate horizons of regular black holes}

\author{Jerzy Matyjasek }
\email{jurek@kft.umcs.lublin.pl}
\affiliation{Institute of Physics,
Maria Curie-Sk\l odowska University\\
pl. Marii Curie-Sk\l odowskiej 1,
20-031 Lublin, Poland}

\author{Pawe{\l} Sadurski}
\affiliation{Institute of Physics,
Maria Curie-Sk\l odowska University\\
pl. Marii Curie-Sk\l odowskiej 1,
20-031 Lublin, Poland}

\author{Dariusz Tryniecki}
\affiliation{{Po\l aniecka 2/69, 22-100 Che\l m}, Poland}
\date{\today}

\begin{abstract}
The regularized stress-energy tensor of the quantized massive scalar, spinor and
vector fields inside the degenerate horizon of the regular charged black hole in the 
(anti-)de Sitter universe is constructed and examined. It is shown that although 
the components of the stress-energy tensor are small in the vicinity of the black 
hole degenerate horizon and near the regular center, they are quite big in the 
intermediate region. The oscillatory character of the stress-energy tensor can 
be ascribed to various responses of the higher curvature terms to the changes 
of the metric inside the (degenerate) event horizon, especially in the region 
adjacent to the region described by the nearly flat metric potentials. Special 
emphasis is put on the stress-energy tensor in the geometries being the product 
of the constant curvature two-dimensional subspaces. 

\end{abstract}
\pacs{04.62.+v,04.70.-s}
\maketitle

\section{Introduction}
The class of the regular black holes, i.e., the black holes for which the
curvature invariants are regular as $r \to 0$ has received much attention
recently. The first constructions of such systems (in which singular interiors
have been replaced by the regular cores) appeared in the mid-1960s in the 
works of Sakharow, Gliner and Bardeen~\cite{Sakharov,Gliner,Bard,Borde,Fernando}. 
Nowadays, quite a number of  solutions of this type are known, supplemented 
by the no-go theorems which forbid their construction in certain 
circumstances~\cite{Bronnikov3}. One of the most interesting and intriguing 
solutions of this type has been proposed by Ayon-Beato and Garcia~\cite{ABG} 
and subsequently reinterpreted by Bronnikov~\cite{Bronnikov2}. It is a static 
and spherically-symmetric solution of the coupled system of nonlinear 
electrodynamics and gravity describing a class of the two-parameter regular 
black holes.  For $r/\mathcal{M} \gg 1$ and for small values of the ratio
$|Q|/\mathcal{M},$ where $Q$ is the (magnetic)  charge and $\mathcal{M}$ 
is the black hole mass, it closely resembles the Reissner-Nordstr\"om solution.  
Noticeable differences appear for the extreme and nearly extreme configurations.
We will refer to this family of the exact solutions of the Einstein equations
as the ABGB black holes.

Although the metric potentials of the $ABGB$ black hole involve the hyperbolic
functions of $r$ it can be demonstrated that the radii of the event and the  
inner horizon can be expressed in terms of the real branches of the Lambert 
functions~\cite{Kocio1,Kocio3}. This property together with the relative 
simplicity of the metric tensor have stimulated continued interest in the ABGB 
black holes~\cite{Radinschi1,Radinschi2,Yang,Jose0,Jose1}.

In this paper  we shall analyze a class of the regular ABGB black holes 
in the asymptotically (anti-)de Sitter universe. It clarifies our earlier 
results presented in Ref.~\cite{klimek} and extends them to the case 
of the negative cosmological constant. Subsequently, we construct 
the stress-energy tensor of the quantized massive scalar, spinor and 
vector field inside the extremal and ultraextremal horizon of the 
degenerate black hole. The calculations are carried out within the framework 
of the Schwinger-De Witt approximation~\cite{fz82,fz83,fz84,kofman1,Jirinek00,Kocio1}. 
This approach is quite general as the sole criterion for its applicability is 
demanding that the ratio of the Compton length associated with the field and 
the characteristic radius of the curvature of the background geometry is small. 
In practice, it turns out that the reasonable results can be obtained for 
$\mathcal{M} m >2$~\cite{PaulA}.

We shall explicitly demonstrate that important and interesting  information 
regarding the stress-energy tensor can easily be obtained by studying 
the geometry of the closest vicinity of the degenerate horizon and the regular 
center. Indeed, near the regular center of the black hole the line
element reduces to that of the (anti-)de Sitter, whereas near the degenerate
event horizon the line element has a product form, where each part describes
maximally symmetric two-dimensional subspace. This is a very fortunate feature 
of the problem as the stress-energy tensor is extremely complicated.  On the 
other hand, the full stress-energy tensor inside the degenerate horizons reveals 
interesting features. Indeed, although the interior of the degenerate regular 
black hole in the asymptotically (anti)-de Sitter universe is described by simple 
functions, the components of the stress-energy tensor in the intermediate region 
are many orders of magnitude greater that the components at the center and in the
neighborhood of the horizon. This behavior, although not quite unexpected, is
totally different from what one usually encounters in the calculations of the
quantum processes in the curved background, provided the calculations
are carried out in the regions that are  sufficiently distant from the central 
singularity. In the regular models the singularity is absent and the oscillatory 
character of the components of the stress-energy tensor is due to  various 
response of the curvature terms constituting $T_{a}^{b}$ to the  changes of 
the metric inside the (degenerate) event horizon. The lengthy and complicated 
formulas describing the components of the stress-energy tensor of the quantized 
massive scalar, spinor and vector fields can be downloaded from the computer 
code repository~\cite{sklad}. Moreover, it should be noted that the general 
solution constructed here provides a natural setting for calculations presented 
in Ref.~\cite{oleg_i_ja}. 
 
The paper is organized as follows:
The regular ABGB black holes in the (anti-)de Sitter universe are introduced
in the next section. In section III the geometries of the vicinity of 
the degenerate horizons are analyzed with the special emphasis put on the
Bertotti-Robinson, Nariai, anti-Nariai and Plebanski-Hacyan solutions. 
The renormalized stress-energy tensor of the massive scalar, spinor and 
vector field inside the horizon of the regular ultraextremal black hole 
is constructed and examined in section IV.
Throughout the paper a natural system of units is adopted. The sign convention
is that of MTW~\cite{MTW}.

\section{Regular black holes in (anti-)de Sitter universe}

The coupled system of equations describing the nonlinear electrodynamics 
and gravity considered in this paper can be constructed form the action 
\begin{equation}
S=\frac{1}{16\pi}\int \left( R-2 \Lambda \right) \sqrt{-g}\,d^{4}x+S_{m},
\end{equation}
where 
\begin{equation}
S_{m}=-\dfrac{1}{16\pi }\int \mathcal{L}\left( F\right) \sqrt{-g}\,d^{4}x
\end{equation}
and
$\mathcal{L}\left( F\right) $ is a functional of $F=F_{ab}F^{ab}$ with $\mathcal{L}(F) \to
F$ as $F \to 0.$ 
All symbols have their usual meaning and   $\Lambda = \varepsilon l^{2}.$ 
The parameter $\varepsilon$  may take one of the three values:  1 for the positive 
cosmological constant,  $-1$ for the negative one and 0 if the cosmological 
term is absent. 

The standard definition of the stress-energy tensor
\begin{equation}
T^{ab}=\frac{2}{\sqrt{-g}}\frac{\delta }{\delta g_{ab}}S_{m}  \label{tensep}
\end{equation}
leads to the following expression
\begin{equation}
T_{a}^{b}=\dfrac{1}{4\pi }\left( \dfrac{d\mathcal{L}\left( F\right) }{dF}%
F_{ca}F^{cb}-\dfrac{1}{4}\delta _{a}^{b}\mathcal{L}\left( F\right) \right) 
\end{equation}
whereas equation of the nonlinear electrodynamics can be written as 
\begin{equation}
 \nabla_{a} \left( \frac{d\mathcal{L}}{dF} F^{ab}\right)\hspace{.5cm} {\rm and} 
 \hspace{.5cm} \nabla_{a}^{\,\,\,*}F^{a b} =0, 
\end{equation}
where an asterix denotes, as usual, the Hodge dual. It is clear that 
the above equations reduce to their classical counterparts as $F \to 0.$

Let us consider spherically-symmetric and static configuration described
by the line element of the form 
\begin{equation}
ds^{2}=-e^{2\psi \left( r\right) }f(r)dt^{2}+\frac{dr^{2}}{f(r)}%
+r^{2}d\Omega ^{2},  
\label{el_gen}
\end{equation}
where $f(r)$ and $\psi(r)$ are unknown functions. 
Since the Lie derivative of the tensor $F_{ab}$ with respect to the generators
of the $O(3)$ group vanish,  the only (independent) components of $F_{ab}$ 
compatible with the assumed  symmetry are $F_{01}$ and $F_{23}$. Simple 
integration yields 
\begin{equation}
F_{23}=Q\sin \theta
\end{equation}
and 
\begin{equation}
r^{2}e^{-2\psi }\dfrac{d\mathcal{L}\left( F\right) }{dF}F_{10}=Q_{e},
\end{equation}
where $Q$ and $Q_{e}$ are the integration constants interpreted as the
magnetic and electric charge, respectively. In the latter we shall assume
that the electric charge vanishes, and, consequently, $F$ is given by 
\begin{equation}
F=\dfrac{2Q^{2}}{r^{4}}.  \label{postacF}
\end{equation}
With the substitution
\begin{equation}
f(r)=1-\frac{2M(r)}{r}
                     \label{fM}
\end{equation}
the time and the radial components of the Einstein field equations
with the cosmological term 
\begin{equation}
G_{a}^{b}+\varepsilon l^{2} \delta_{a}^{b} = 8 \pi T_{a}^{b}
\label{eins}
\end{equation}
assume simple and transparent form
\begin{equation}
-\frac{2}{r^{2}}\frac{dM}{dr}+\varepsilon l^{2} = -\frac{1}{2} \mathcal{L}(F) 
\end{equation}
and
\begin{equation}
-\frac{2}{r^{2}}\frac{dM}{dr}+\frac{2}{r}\left( 1-\frac{2M}{r}\right) \frac{d\psi }{dr}
+ \varepsilon l^{2} = -\frac{1}{2} \mathcal{L}(F)  \label{1st}
\end{equation}
and can  be formally integrated.
The angular component of (\ref{eins}) 
\begin{equation}
\left( 1-\frac{2 M}{r}\right)\left[\frac{d^{2}\psi}{dr^{2}} +  
\left(\frac{d\psi}{dr}\right)^{2}\right]
-\frac{M}{r} \frac{d^{2} M}{dr^{2}} + \frac{1}{r}\left(\frac{M}{r} 
- 3 \frac{dM}{dr} +1\right) \frac{d\psi}{dr}
+ \varepsilon l^{2} = 2\dfrac{d\mathcal{L}
\left( F\right) }{dF}\dfrac{Q^{2}}{r^{4}}-\dfrac{1}{2 }\mathcal{L}\left(F\right)
\end{equation}
is merely constraint equation.
Further considerations require specification of the Lagrangian $\mathcal{L}
\left( F\right) .$ 
We demand that it should have proper asymptotic, i.e., in a weak field limit it 
should approach
$F.$
Following Ay\'on-Beato, Garc\'\i a~\cite{ABG} and Bronnikov~\cite{Bronnikov2} 
let us
chose it in the form 
\begin{equation}
\mathcal{L}\left( F\right) \,=F\left[ 1-\tanh ^{2}\left( s\,\sqrt[4]{\frac{
Q^{2}F}{2}}\right) \right] ,  \label{labg}
\end{equation}
where 
\begin{equation}
s=\frac{\left| Q\right| }{2b},  \label{sabg}
\end{equation}
and the free parameter $b$ will be adjusted to guarantee regularity at the
center.
Inserting Eq.~(\ref{sabg}) into (\ref{labg}) and making use of Eq.~(\ref
{postacF}) one has 
\begin{equation}
8\pi T_{t}^{t}=8\pi T_{r}^{r}=-\frac{Q^{2}}{r^{4}}\left( 1-\tanh ^{2}\frac{
Q^{2}}{2br}\right) .  \label{tep}
\end{equation}
Now the equations can easily be integrated in terms of the elementary functions:
\begin{equation}
M\left( r\right) =C_{1}-b\tanh \frac{Q^{2}}{2br}+\frac{\varepsilon l^{2}  r^{3}}{6},
\hspace{0.4cm}\psi \left( r\right) =C_{2}  \label{mm0}
\end{equation}
where $C_{1}$ and $C_{2}$ are the integration constant. Making use of the
condition 
\begin{equation}
  \psi(\infty)=0
\end{equation}
gives  $C_{2} = 0.$ 
The next step requires some prescience: let us assume that the solution describes 
black hole and its event horizon is located at $r = r_{+}.$ The integration 
constant 
$C_{1}$ can be determined form the condition 
\begin{equation}
M(r_{+}) = \frac{r_{+}}{2}.
\end{equation}
The solution for $M(r)$ can be written in the form
\begin{equation}
M(r) = \frac{r_{+}}{2} + b \tanh\frac{Q^{2}}{2 b r_{+}} -\frac{\varepsilon l^{2} r_{+}^{3}}{6} -
                       b \tanh\frac{Q^{2}}{2 b r} + \frac{\varepsilon l^{2} r^{3}}{6}.
\end{equation}
It can be demonstrated that for $\varepsilon = -1$ the first three terms in the 
right hand side of the above equation comprise the Abbott-Deser mass of the black
hole:
\begin{equation}
\mathcal{M}_{AD} = \frac{r_{+}}{2} + b \tanh\frac{Q^{2}}{2 b r_{+}} +\frac{ l^{2} r_{+}^{3}}{6}.
\label{Madm}
\end{equation}
On the other hand the $\varepsilon = 1$ case is slightly more subtle. Nevertheless,
one can always refer to the horizon-defined mass~\footnote{In Ref.~\cite{klimek} 
the argumentation leading to relation of the integration constant to the black 
hole mass is erroneous. The resulting line element is, however, correct.}
\begin{equation}
\mathcal{M}_{H} =  \frac{r_{+}}{2} + b \tanh\frac{Q^{2}}{2 b r_{+}} - \frac{ l^{2} r_{+}^{3}}{6}.
\label{Mhor}
\end{equation}
In the latter, for brevity, we shall 
denote both masses by a single symbol $\mathcal{M}.$
Demanding the regularity of the line element as $r\rightarrow 0$ yields 
$b = C_{1} \equiv  \mathcal{M,}$ and, consequently, the resulting line element has 
the form (\ref{el_gen}) with $\psi(r)=0$ and
\begin{equation}
f(r) = 1-\frac{2 \mathcal{M}}{r}\left(1-\tanh\frac{Q^{2}}{2 \mathcal{M} r}
\right)-\frac{\varepsilon l^{2} r^{2}}{3}.
        \label{el_gen1}
\end{equation}
It should be noted that with such a choice of $C_{1}$ and $b$ both (\ref{Madm}) 
and (\ref{Mhor})  are  consistent with (\ref{el_gen1}), i.e., depending on the sign
of the cosmological constant $f(\rp)=0$ is 
equivalent either to (\ref{Madm}) or to (\ref{Mhor}).  
We shall call this solution the Ay\'on-Beato-Garc\'\i a-Bronnikov-(anti-)de Sitter 
solution (ABGB-(a)dS). It could be easily shown that putting $Q=0$ yields 
the Schwarzschild-de Sitter (Kottler) solution, whereas for $\Lambda=0$ one 
gets the Ay\'on-Beato, Garc\'\i a  line element as reinterpreted by Bronnikov.

\begin{figure}[h]
\includegraphics[width=11cm]{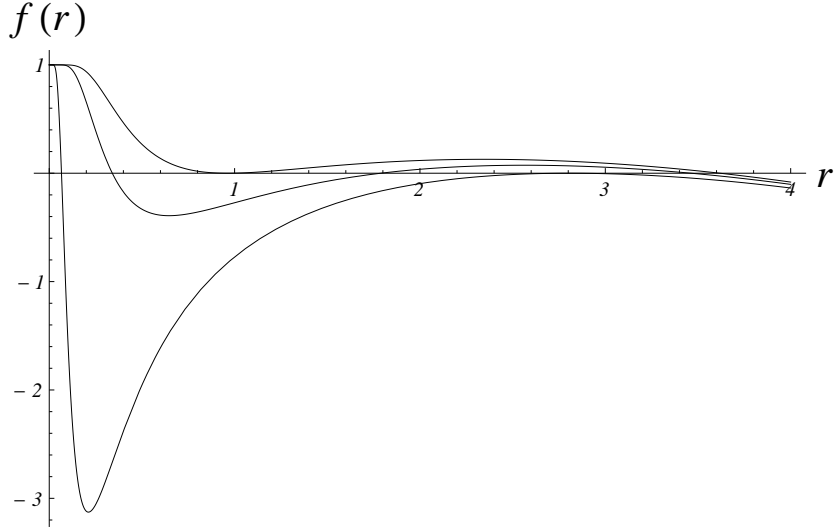}
\caption{The qualitative behavior of the function $f(r)$ for 
$r_{-} =r_{+} <r_{c},$ $r_{-} < r_{+} < r_{c}$ and $r_{c} < r_{+} = r_{c}.$    }
\label{pp}
\end{figure}

To study ABGB-(a)dS line element it is convenient to introduce the
dimensionless quantities $x=r/\mathcal{M}$, $q=\left| Q\right| /\mathcal{M}$ 
and $\ell=l M$. Here we shall concentrate on configurations with at least 
one horizon. First, let us observe that for small charges 
$\left(q\ll 1\right) $ as well as at great distances form 
the black hole $\left(x\gg 1\right) $ the ABGB-(a)dS solution closely 
resembles that of RN-(a)dS. 
Indeed, expanding the function $f\left( r\right)$ 
in powers of $q^{2}$ (or in powers of $x^{-1}$) one obtains
\begin{equation}
f\,=1-\frac{2\mathcal{M}}{r}+\frac{Q^{2}}{r^{2}}\,
-\frac{\varepsilon l^{2} r^{2}}{3}-\,
\frac{Q^{6}}{12\mathcal{M}^{2}r^{4}}\,+\,....
\end{equation}
Similarly, near the center one has
\begin{equation}
f\sim 1-\frac{4\mathcal{M}}{r}\exp \left( \frac{-Q^{2}}{\mathcal{M}r}\right)
-\frac{\varepsilon l^{2} r^{2}}{3}
\end{equation}
and the metric in the closest vicinity of $r=0$ may by approximated by the
(anti-)de Sitter line element as the second term in the right hand side of 
the approximation rapidly goes to zero. It should be noted that for a given 
$\mathcal{M}$ and $Q$ and  small cosmological constant ($\ell^{2} \ll 1 $)
some of the features of the ABGB-(a)dS geometry are to certain extend  
similar to that of the  ABGB spacetime. Simple analysis shows that for 
$\varepsilon =1$ there are, at most, three distinct positive roots of the 
equation $f\left( r\right) =0.$  One expects that the two of them are 
located closely to the inner and event horizons of the ABGB black hole, 
whereas the third one (absent in the ABGB geometry) is  approximately located 
at $x_{c}\simeq \sqrt{3}/\ell$  and interpreted as the cosmological horizon.  
For $\varepsilon =-1$ these similarities are even more transparent as 
there is no the cosmological horizon. Qualitative behavior of the function $f(r)$ 
with $\varepsilon >0$ is displayed in Fig.~\ref{pp}.

\begin{figure}[h]
\includegraphics[width=11cm]{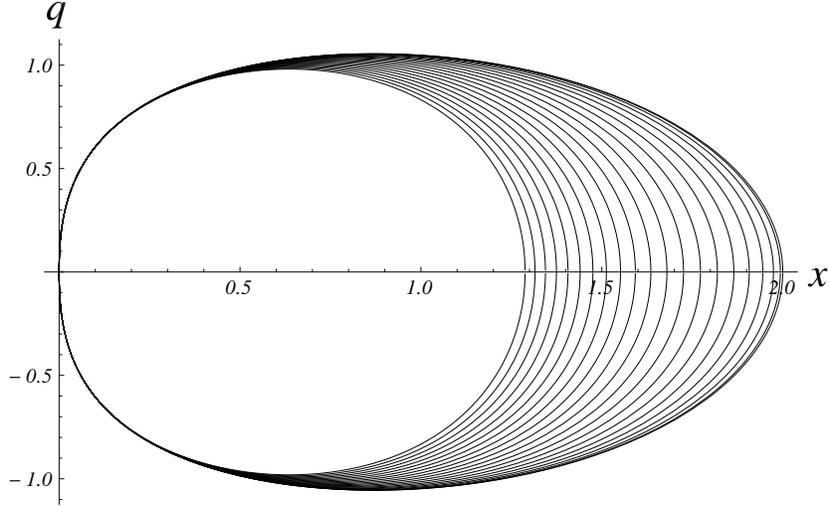}
\caption{The charge $q$ plotted as function of $x.$ Each curve is labeled 
by the negative cosmological constant. The curves are drawn for  $\ell = 0.05 i$ 
( $i =0,..., 20$). The outermost curve represents the ABGB black hole. 
The extrema of the curves represent degenerate configurations.}
\label{rys1}
\end{figure}

\begin{figure}[h]
\includegraphics[width=11cm]{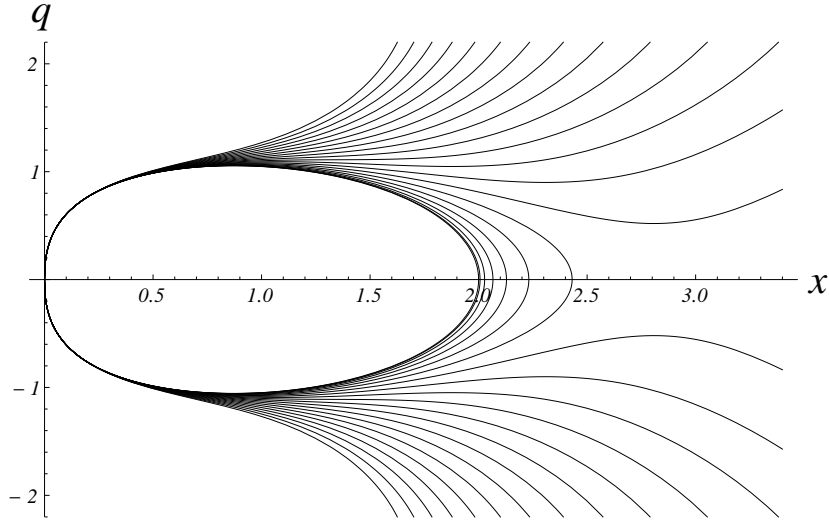}
\caption{The charge $q$ plotted as function of $x.$ Each curve is labeled by  
positive cosmological constant. The curves are drawn for $\ell = 0.05 i$ 
($i =0,..., 20$). The innermost curve represents the ABGB black hole. 
The extrema of the curves represent degenerate configurations.}
\label{rys2}
\end{figure}

Although it is impossible for $\varepsilon \neq 0$ to give exact solutions 
representing location of the horizons, one can easily solve
\begin{equation}
1-\frac{2}{x}\left( 1-\tanh\frac{q^2}{2x}\right) 
-\frac{1}{3}\varepsilon \ell^{2} x^2 =0
                                       \label{eqq}
\end{equation}
with respect to $q.$ Simple manipulations give
\begin{equation}
q = \pm \sqrt{x \ln \frac{12-3x+\varepsilon \ell^{2} x^{3}}{x(3-\varepsilon \ell^{2} x^{2})}}.
\end{equation}
The function $q(x)$  is plotted in Figs.~\ref{rys1} and~\ref{rys2}, where each 
curve is labeled by the cosmological constant $\varepsilon \ell.$  The inner, 
event and cosmological horizon (denoted by $x_{-},$ $x_{+}$ 
and $x_{c},$ respectively) lie on intersections of the $q-$constant line
and the curve drawn for a constant $\ell.$ For $\varepsilon \leq 0$ the 
outermost curve represent the special case of the ABGB black hole with 
the inner and event horizon expressed in terms of the real branches of the 
Lambert W function~\cite{Kocio1,Kocio3}. The extrema of each curve represent 
degenerate configuration with $x_{-} =x_{+}.$  Especially interesting is 
the  solution describing the extreme ABGB black hole~\cite{Kocio3,Kim}. 

For $\varepsilon >0$ 
Eq.~(\ref{eqq}) has, in general, three positive roots, which can merge leading 
to various interesting configurations. Indeed, for special choices of the 
parameters one can have a configuration with a degenerate and a nondegenerate 
horizon or one triply degenerate horizon. The first configuration is characterized 
by $r_{-}<r_{+}<r_{c}$ whereas the second configuration contains two subclasses 
depending on which horizons do merge. As the degenerate horizons are located 
at simultaneous zeros of 
$f\left( r\right) $ and $f^{\prime }\left(r\right)$ 
they differ by a sign of the second derivative of $f.$ The first of degenerate 
configurations, called cold black hole, is characterized by $f''(r_{+})>0$ 
and  $r_{-}=r_{+}<r_{c}.$  On the other hand, the configuration
characterized by $f''(r_{+}) <0$ and $r_{-}<r_{+}=r_{c},$ is usually referred 
to as the charged Nariai black hole.  Finally, for the triply degenerate horizon 
characterized by $r_{-}=r_{+}=r_{c}$ occurs for $f''(r_{+}) =0.$  This 
configuration is characterized by $q_{crit}=1.1082,$   $x_{crit}=1.34657$ 
and $\ell_{crit} = 0.496.$ 
Note that regardless of the sign of the cosmological constant the radial
coordinate of the inner horizon, $x_{-},$ for $|q| \lesssim 0.9$ is weakly influenced by $\ell$
and is close to its $\ell=0$ value. 

For $\varepsilon >1$  one can single out the lukewarm black hole, for which the
surface gravity of the event horizon equals the surface gravity of the cosmological
horizon. Since the Hawking temperatures of the horizons are equal,
one can construct a regular thermal state. The lukewarm Reissner-Nordstr\'om-de Sitter
black holes have been extensively studied in 
Refs.~\cite{Romans,Cai,Winstanley,Breen2010,Matyjasek2a,Matyjasek2,Breen2011,Sadurski}.  

The Penrose diagrams representing a two-dimensional $t-r$ section of the conformally 
transformed ABGB-(a)dS geometry are in many respects similar to 
the analogous diagrams constructed for the Reissner-Nordstr\"om-(anti-)de Sitter 
black holes~\cite{Lake,Mellora,Mellor,Banados} with central singularity replaced by  
a regular region. Indeed, for $\varepsilon =1$ one has a two-dimensional infinite 
carpet, which can be obtained by vertical and horizontal translations of the simple
diagram representing two black holes in asymptotically de Sitter universe.
For $\varepsilon =0$ the Penrose diagram is similar to the diagram drawn for
the Reissner-Nordstr\"om black hole. Finally, for $\varepsilon= -1$ the diagram
is still similar to the Reissne-Nordstr\"om case with the conformal null infinity
replaced by the anti-de Sitter infinity.

\section{The geometries of the closest vicinity of the degenerate horizons}

The curvature scalar for the line element (\ref{el_gen})  with $\psi(r) = 0$ 
is given by
\begin{equation}
R = -\frac{d^{2}f}{dr^{2}} -\frac{4}{r} \frac{df}{dr} -\frac{2f}{r^{2}} + \frac{2}{r^{2}}.
\end{equation}
At the degenerate horizon both $f(r_{+})$ and $f'(r_{+})$ vanish and the Ricci 
scalar there is given by
\begin{equation}
R = -{\frac{d^{2}f}{dr^{2}}}_{|r_{+}} + \frac{2}{r_{+}^{2}}
\end{equation}
whereas at the degenerate horizon of ultraextemal black hole $R =2/r_{+}^{2}.$
This behavior of the curvature scalar is a manifestation of the fact that the 
geometry of the closest vicinity of the degenerate horizons when expanded
into a whole manifold is a direct product of the two maximally symmetric
2-dimensional subspaces. 
It can be demonstrated as follows: for the nearly extreme
configuration the function $f$ can be approximated by a parabola
\begin{equation}
f(r) =\frac{1}{2}f''(r_{d})(r-r_{1})(r-r_{2}),
\end{equation}
where $x_{d}$ represents a degenerate horizon, and $r_{1}$ and $r_{2}$ denote 
a pair of close horizons, i.e., either $r_{-}$ and $r_{+}$ or $r_{+}$ and $r_{c}.$
For $f''(r_{d}) >0$ one can introduce new coordinates $t=2 T/(\Delta f''(r_{d})$ and
$r =r_{d} + \Delta \cosh y$ and take the limit $\Delta \to 0$
to obtain
\begin{equation}
 ds^{2} = \frac{2}{f''(r_{d})}\left(-\sinh^{2}y\, dT^{2} +dy^{2}\right) 
+ r^{2}_{d} \,d \Omega^{2}.
\label{bebe}
\end{equation}
Similarly, for $f''(r_{d}) <0$  the transformation given by 
$t=2 T/(\Delta f''(r_{d}))$ and $r =r_{d} + \Delta \cos y$ yields 
\begin{equation}
ds^{2} = \frac{2}{f''(r_{d})}\left(\sin^{2}y \,dT^{2} - dy^{2}\right) + r^{2}_{d} \,
d \Omega^{2}.
\label{hidekazu}
\end{equation}
Finally, for the ultraextremal black hole, putting $y= \eta \sqrt{f''(r_{d})/2} $ and
taking limit $f''(r_{d}) \to 0$ one obtains
\begin{equation}
 ds^{2} = - \eta^{2} dT^{2} +d\eta^{2} + r_{d}^2 \left(d\theta^{2} + \sin^{2} \theta d\phi^{2}\right).
 \label{pleban}
\end{equation}

Although there are no commonly accepted  names for line elements 
(\ref{bebe}-\ref{pleban}) it seems that the convention of Podolsky and 
Griffiths~\cite{Podolsky}  is the most appropriate in this regard. Consequently 
we shall address to (\ref{bebe}) and (\ref{hidekazu})  as the Bertotti-Robinson 
and Nariai line element, respectively, even though the modulus of the curvature 
radii of the 2-dimensional maximally symmetric subspaces are nonequal. 
The line element (\ref{pleban}) will be addressed to as the Pleba\'nski-Hacyan 
solution. All the cases considered above are tabulated in Table I. 
It should be noted that in general, the family of geometries being the  product 
of the constant curvature spaces is richer and allows for two-dimensional 
Euclidean ($E^{2}$) and hyperbolic ($H^{2}$) spaces.
\begin{table}[ht]
\begin{tabular}{|c|c|r|r|}
\toprule
$\varepsilon$ &$f'(r_{+})$ &$f''(r_{+})$ &topology
\\ \hline
1 &0  &$>0$ & $AdS_{2}\times S^{2}$
\\ \hline
1 &0  &$<0$ & $dS_{2}\times S^{2}$
\\ \hline
1 &0 & 0 & $M_{2}\times S^{2}$
\\ \hline
-1 & 0 & $>0$ & $AdS_{2}\times S^{2}$
\\ \hline
0 &0& $ >0$ & $AdS_{2}\times S^{2}$
\\
\botrule
\end{tabular}
\caption{The geometries of the closest vicinity of the degenerate event 
horizon. $\varepsilon$ gives the sign of the cosmological constant. 
The configurations with $f''(r_{+})>0$ are described by the Bertotti-Robinson 
line element, the configurations with $f''(r_{+})<0$ are described by the 
Nariai line element. The configuration with vanishing second derivative is 
described by the Pleba\'nski-Hacyan geometry.}
\end{table}

\section{Renormalized stress-energy tensor}
The approximate stress-energy tensor employed here is constructed from the 
one-loop effective action 
 \begin{eqnarray}
 W^{(1)}_{ren}\,&=& \,{1\over 192 \pi^{2} m^{2}} \int d^{4}x g^{1/2}
 \left( \al^{(s)}_{1} R \Box R
+\al^{(s)}_{2} R_{a b} \Box R^{a b}
+ \al^{(s)}_{3} R^{3}
+\al^{(s)}_{4} R R_{a b} R^{a b} 
\right. \nonumber \\
&&
+\al^{(s)}_{5} R R_{a b c d} R^{a b c d}        
+\al^{(s)}_{6} R^{\pp{b} a}_{b} R^{\pp{c} b}_{c} R^{\pp{a} c}_{a}
+\al^{(s)}_{7} R_{a b} R^{c d} R_{c \pp{a} d}^{\pp{c} a \pp{d}  b}
+\al^{(s)}_{8} R_{a b} R^{a}_{\pp{a} e c d} R^{b e c d}
\nonumber \\
&&\left. 
+\al^{(s)}_{9} R_{c d}^{\pp{c} \pp{d} a b} R_{a b}^{\pp{a} \pp{b} e h}
R_{ e h}^{\pp{e} \pp{h} c d}
+\al^{(s)}_{10} R_{c ~ d}^{~ a ~ b} R_{a ~ b}^{~ e ~ f} R_{e ~ f}^{~ c ~ d}
 \right),
\label{dzialanie}
\end{eqnarray}
where $m$ is the mass of the field and the numerical coefficients depending on 
the spin of the field are  given in a Table~\ref{table1}.
\begin{table}[h]
\begin{tabular}{|c|c|c|c|}\hline
& s = 0 & s = 1/2 & s = 1 \\\hline\hline
$\al^{(s)}_{1} $ & $ {1\over2}\xi^{2}\,-\,{1\over 5} \xi $\,+\,${1\over
56}$ & $- {3\over 280}$ &
$- {27\over 280}$\\
 $\al^{(s)}_{2}$ & ${1\over 140}$ & ${1\over 28}$ & ${9 \over 28}$\\
 $\al^{(s)}_{3}$ &$ \left( {1\over 6} - \xi\right)^{3}$& ${1\over 864}$ &$
- {5\over 72}$\\
 $\al^{(s)}_{4}$ & $- {1\over 30}\left( {1\over 6} - \xi\right) $& $-
{1\over 180}$ & ${31\over 60}$\\

 $\al^{(s)}_{5}$ & ${1\over 30}\left( {1\over 6} - \xi\right)$ &$ -
{7\over 1440}$ &$ - {1\over 10}$\\

 $\al^{(s)}_{6}$ & $ - {8\over 945}  $& $- {25 \over 756}$ & $- {52\over
63}$\\

 $\al^{(s)}_{7}$ & ${2 \over 315}$ & $ {47\over 1260}$  & $- {19\over 105} $\\
 $\al^{(s)}_{8}$ & ${1\over 1260}$ & ${19\over 1260} $ & ${61\over  140} $\\
 $\al^{(s)}_{9}$ & ${17\over 7560}$& ${29\over 7560}$ & $- {67\over 2520}$\\
 $\al^{(s)}_{10}$ & $- {1\over 270}$ & $ - {1\over 108} $  & $ {1\over 18}$\\ \hline
 \end{tabular}
 \caption{The coefficients $\al_{i}^{(s)}$ for the massive scalar with arbitrary
curvature coupling $\xi$ , spinor,
and vector
 field}
 \label{table1}
 \end{table}
The tensor 
\begin{equation}
T^{ab} = \frac{2}{\sqrt{g}} \frac{\delta  W_{ren}^{(1)}}{\delta g_{ab}}
\label{set}
\end{equation}
is known to yield reasonable results so long the length of the Compton wave 
associated with the field is smaller than the characteristic radius of curvature 
of the background geometry. This is satisfied in a number of physically 
interesting cases and allows for weak temporary changes. 
The most general expression constructed by functional differentiations
of the effective action with respect to the metric tensor has been
constructed in Refs.~\cite{Jirinek00,Kocio1}, to which the reader
is addressed for computational details. The approximate stress-energy 
tensor consists of almost 100 purely geometric terms constructed from
the curvature tensor and its covariant derivatives. For $N$ fields $\psi_{i}$ 
of spin $s$ characterized by (possibly various) masses $m_{i},$ 
the one-loop effective action is still of the form (\ref{dzialanie})  with 
\begin{equation}
\frac{1}{m^{2}} \to \sum_{i=1}^{N} \frac{1}{m_{i}^{2}}.
\end{equation}
Consequently, the quantum effects can be made arbitrary large simply 
by taking a large number of fields into account. For the quantized massive 
scalar field with the arbitrary curvature coupling  in the spacetime 
of spherically-symmetric and asymptotically flat, static black holes  
the tensor (\ref{set}) coincides with the tensor constructed in Ref.~\cite{AHS}. 

Equipped with the results of the previous section one can easily calculate the
renormalized stress-energy tensor of the massive quantized fields in the geometries,
tabulated in Table I. Making use of the general formulas presented in 
Refs.~\cite{Jirinek00,Kocio1} one can show that in the vicinity of the degenerate 
horizon the components of the stress-energy tensor of the massive scalar, 
spinor and vector field for the line element 
(\ref{el_gen}) with $\psi(r) =0$  reduce to 
\begin{equation}
 \bar{T}_{t}^{(q)t} = \bar{T}_{r}^{(q)r} = s_{1}^{(i)} \frac{(f''(r_{d}))^{2}}{2 r_{+}^{2}} 
 + s_{2}^{(i)} (f''(r_{d}))^{3} + 4 s_{2}^{(i)}\frac{1}{r_{+}^{6}}
 \label{sub1}
\end{equation}
and
\begin{equation}
 \bar{T}_{\theta}^{(q)\theta} = \bar{T}_{\phi}^{(q)\phi} = 
 -s_{1}^{(i)} \frac{f''(r_{d})}{ r_{+}^{4}} 
 -\frac{1}{2} s_{2}^{(i)} (f''(r_{d}))^{3} - 8 s_{2}^{(i)}\frac{1}{r_{+}^{6}},
 \label{sub2}
\end{equation}
where $\bar{T}_{a}^{(q)b} = 96 \pi^{2} m^{2} T_{a}^{(q)b}$ and the coefficients 
$s_{i}^{(j)}$ for scalar, spinor and vector fields are listed in Table II.
\begin{table}[ht]
\begin{tabular}{|c|c|c|c|}
\toprule
$s_{i}^{(j)}$ & $j=0$ & $j=\frac{1}{2}$ & $j=1$
\\ \hline
$i=1$  &$-\frac{1}{30} + \frac{8}{15}\xi -3 \xi^{2} + 6 \xi^{3} $& $ \frac{1}{120}$ &
$\frac{1}{10}$
\\ \hline
$i=2$ &$\frac{1}{105} -\frac{1}{10}\xi +\frac{1}{2}\xi^{2} -\xi^{3}$ &$\frac{1}{168}$ & 
$\frac{1}{35}$
\\
\botrule
\end{tabular}
\caption{The spin-dependent numerical coefficients standing in front 
of the geometric terms in Eqs. (\ref{sub1}) and (\ref{sub2}). }
\end{table}
Inspection of the stress-energy tensor shows that it depends only 
on the curvature radii of the maximally symmetric subspaces, as expected. 
The type of the field enters the equations through spin-dependent numerical 
coefficients. This result can easily be generalized to all spaces with 
symmetric 2-dimensional subspaces. The general line element describing 
nine possibilities of product manifolds (six of which is physical) can be 
written in the compact form
\begin{equation}
ds^{2} = - \frac{2 du dv}{\left(1
- \frac{1}{2} \varepsilon_{1} u v a^{- 2}\right)^{2}}
+  \frac{2 d\zeta d\bar{\zeta}}{\left(1+ \frac{1}{2}\varepsilon_{2} \zeta 
\bar{\zeta}{b^{-2}}\right)^{2}}
\end{equation}
where $a$ and $b$ are related to the Gaussian curvature 
$K_{1} =\varepsilon_{1} a^{-2}$ and $K_{1} =\varepsilon_{2} b^{-2},$ 
respectively, and both $\varepsilon_{1}$ and $\varepsilon_{2}$
can take three values -1,0,1. All physical geometries are displayed in 
Table IV~\cite{Podolsky}.

\begin{table}[ht]
\begin{tabular}{|c|c|r|r|}
\toprule
manifold &topology &$\varepsilon_{1}$ &$\varepsilon_{2}$ 
\\ \hline
Minkowski &$M_{2}\times E^{2}$ &0 &0
\\ \hline
Pleba\'nski-Hacyan & $M_{2} \times S^{2}$ &0 &1
\\ \hline
Pleba\'nski-Hacyan & $AdS_{2}\times E^{2}$ &-1 &0
\\ \hline
Nariai & $dS_{2}\times S^{2}$ & 1 & 1
\\ \hline
Bertotti-Robinson &$AdS_{2}\times S^{2}$ &-1 &1
\\ \hline
anti-Nariai &$AdS_{2}\times H^{2}$ &-1 &-1
\\
\botrule
\end{tabular}
\caption{The product geometries with maximally symmetric two-dimensional 
subspaces.}
\end{table}
The nonzero components of the renormalized stress-energy tensor for all 
six geometries can be written compactly as
\begin{eqnarray}
\bar{T}_{u}^{u} =\bar{T}_{v}^{v} &=&
-8\,{\frac {{{\it \varepsilon_{1}}}^{2}{\it \varepsilon_{2}}\,
{\it c_{3}}}{{b}^{2}{a}^{4}}}
-2\,{\frac { \left( 2\,{\it \varepsilon_{1}}{b}^{6}+{{\it 
\varepsilon_{1}}}^{2}{b}^{4}{\it \varepsilon_{2}}\,{a}^{2}
-{\it \varepsilon_{2}}{a}^{6} \right) 
\left( 2\,{\it c_{3}}+{\it c_{4}}
+2\,{\it c_{5}} \right) }{{b}^{6}{a}^{6}}} 
\nonumber \\
&&-
{\frac { \left( 2\,{\it \varepsilon_{1}}{b}^{6}
-{\it \varepsilon_{2}}{a}^{6} \right)  \left( {\it c_{6}}
+{\it c_{7}}+2\,{\it c_{8}}
+4\,{\it c_{9}} \right) }{{b}^{6}{a}^{6}}}
\label{prod1}
\end{eqnarray}
and
\begin{eqnarray}
\bar{T}_{\theta}^{\theta} =\bar{T}_{\phi}^{\phi} &=&-8\,
{\frac {{{\it \varepsilon_{2}}}^{2}{\it \varepsilon_{1}}\,
{\it c_{3}}}{{a}^{2}{b}^{4}}}
+ 2\,{\frac { \left( {\it \varepsilon_{1}}{b}^{6}-{{\it \varepsilon_{2}}}
^{2}{\it \varepsilon_{1}}\,{a}^{4}{b}^{2}-2\,{\it \varepsilon_{2}}{a}^{6}
 \right)  \left( 2\,{\it c_{3}}+{\it c_{4}}+2\,{\it c_{5}} \right) }{{b}^{6}
{a}^{6}}}\nonumber \\
&&
+{\frac { \left( {\it \varepsilon_{1}}{b}^{6}-2\,{\it 
\varepsilon_{2}}{a}^{6} \right)  \left( {\it c_{6}}+{\it c_{7}}+2\,{\it c_{8}}
+4\,{\it c_{9}} \right) }{{b}^{6}{a}^{6}}},
\label{prod2}
\end{eqnarray}
where, for typographical reasons, we put $c_{i} = \alpha^{(s)}_{i}.$
It can be  shown that the tensor (\ref{prod1}) and (\ref{prod2}) 
calculated for the Bertotti-Robinson, Nariai and Pleba\'nski-Hacyan 
$(M_{2} \times S^{2})$ geometries reduces to (\ref{sub1}) and (\ref{sub2}).
Indeed, simple manipulations give 
\begin{equation}
a^{2} = \frac{2\varepsilon_{3}}{|f''(\rp)|} \hspace{3mm} {\rm and} \hspace{3mm} b=\rp,
\end{equation} 
where $\varepsilon_{3} =-1$ for the Nariai geometry and $\varepsilon_{3} =1$
for the Bertotti-Robinson geometry.

Before going further let us calculate the renormalized stress-energy tensor in
the closest vicinity of the regular center of the ABGB-(a)dS black hole. 
One expects, that the result will depend on solely on the cosmological constant 
as the  de Sitter and  Anti-de Sitter spacetimes are maximally symmetric. 
Making use of the general expressions one has
\begin{equation}
 \bar{\mathfrak{T}}_{a}^{(i)b} = -\frac{1}{9} \varepsilon l^{6} \left(144 c_{3} 
+ 36 c_{4} + 24 c_{5} + 9 c_{6} 
+ 9 c_{7} + 6 c_{8} + 4 c_{9} + 2 c_{10}\right) {\rm diag}[1,1,1,1]_{a}^{b}, 
\label{deS}
\end{equation}
where the expression in parentheses  reduces to 
\begin{equation}
 \frac{1}{315}\left(185-3654\xi+22680\xi^{2}-45360\xi^{3}\right)
\end{equation}
for the masive scalars,
\begin{equation}
 -\frac{31}{1260}
\end{equation}
for massive spinors
and
\begin{equation}
 -\frac{5}{21}
\end{equation}
for vectors.
One expects that it is the stress-energy tensor of the quantized
massive fields in the closest vicinity of the black hole center.

Now  we shall construct the stress-energy tensor of the quantized
massive fields inside the extremal black hole. The calculations may by thought
of as generalization of the analogous calculation carried out in Ref.~\cite{AHL}.
There are, however, notable differences: The geometry under consideration
is regular and static.

To analyze the general stress-energy tensor in the black hole interior it 
is helpful to introduce two functions
$\beta(r) = 1-\tanh(Q^{2}/2\mathcal{M}r) $ and 
$\omega(r) = 1+\tanh(Q^{2}/2 \mathcal{M}r). $
It could be easily checked that their derivatives can be expressed in terms 
of themselves and powers of $r^{-1}.$
The stress-energy tensor expressed in terms of the dimensionless $x, q$ and $\ell$
 has the general structure
\begin{equation}
\bar{T}_{a}^{(i)b} =\frac{1}{\mathcal{M}^{6}} \sum_{n=1}^{15}\sum_{p=1}^{6}\sum_{s=1}^{6}\sum_{t=0}^{2}
\sum_{u=1}^{8} [\alpha^{(i)}_{npstu}]_{a}^{b}
\frac{q^{2s} \ell^{2t}}{x^{n}}\omega^{p}(x)\beta^{u}(x)
+ \bar{\mathfrak{T}}_{a}^{(i)b}, 
\end{equation}
where the first term in the right hand side vanishes as $r \to 0$
and the second approaches the stress-energy tensor of the 
quantized massive field in the (anti-)deSitter spacetime.
The difference between the radial and time components of the stress-energy
tensor factors as
\begin{equation}
\bar{T}_{r}^{(i)r} - \bar{T}_{t}^{(i)t} = g_{tt} F(r),
\end{equation}
where $F(r)$ is a regular function with $F(0) =0$ and hence the stress-energy
tensor is regular in the physical sense.

\begin{figure}[h]
\includegraphics[width=11cm]{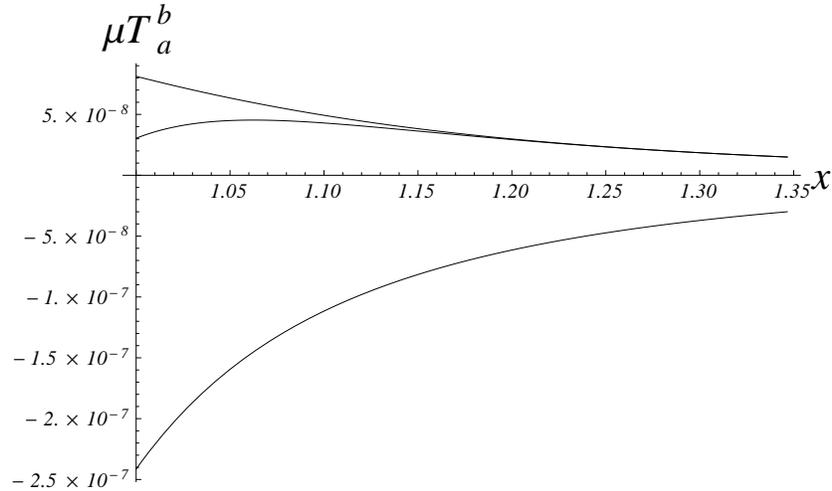}
\caption{The rescaled components of the stress-energy tensor of the conformally coupled
massive scalar field in the vicinity of the ultraextremal horizon of the regular black hole.
Top to bottom the curves represent the radial, time and angular components, respectively.
$[\mu =\mathcal{M}^{6}m^{2}/10^{2}]$.}
\label{pp2a}
\end{figure}

\begin{figure}[h]
\includegraphics[width=11cm]{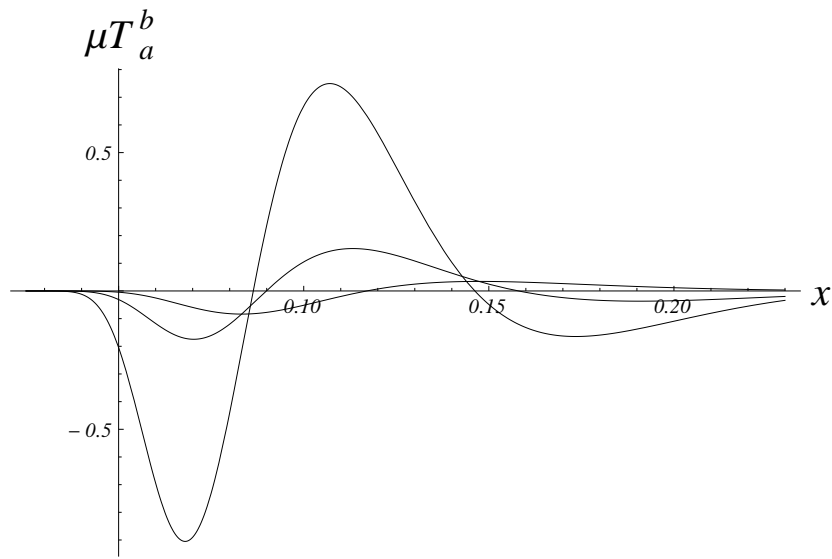}
\caption{The rescaled components of the stress-energy tensor of the conformally
coupled massive scalar field inside the ultraextremal regular black hole. Top to 
bottom at $x=0,1$  the curves represent the time, angular and radial components
of the stress-energy tensor. $[\mu =\mathcal{M}^{6}m^{2}/10^{2}]$. }
\label{pp2b}
\end{figure}

\begin{figure}[h]
\includegraphics[width=11cm]{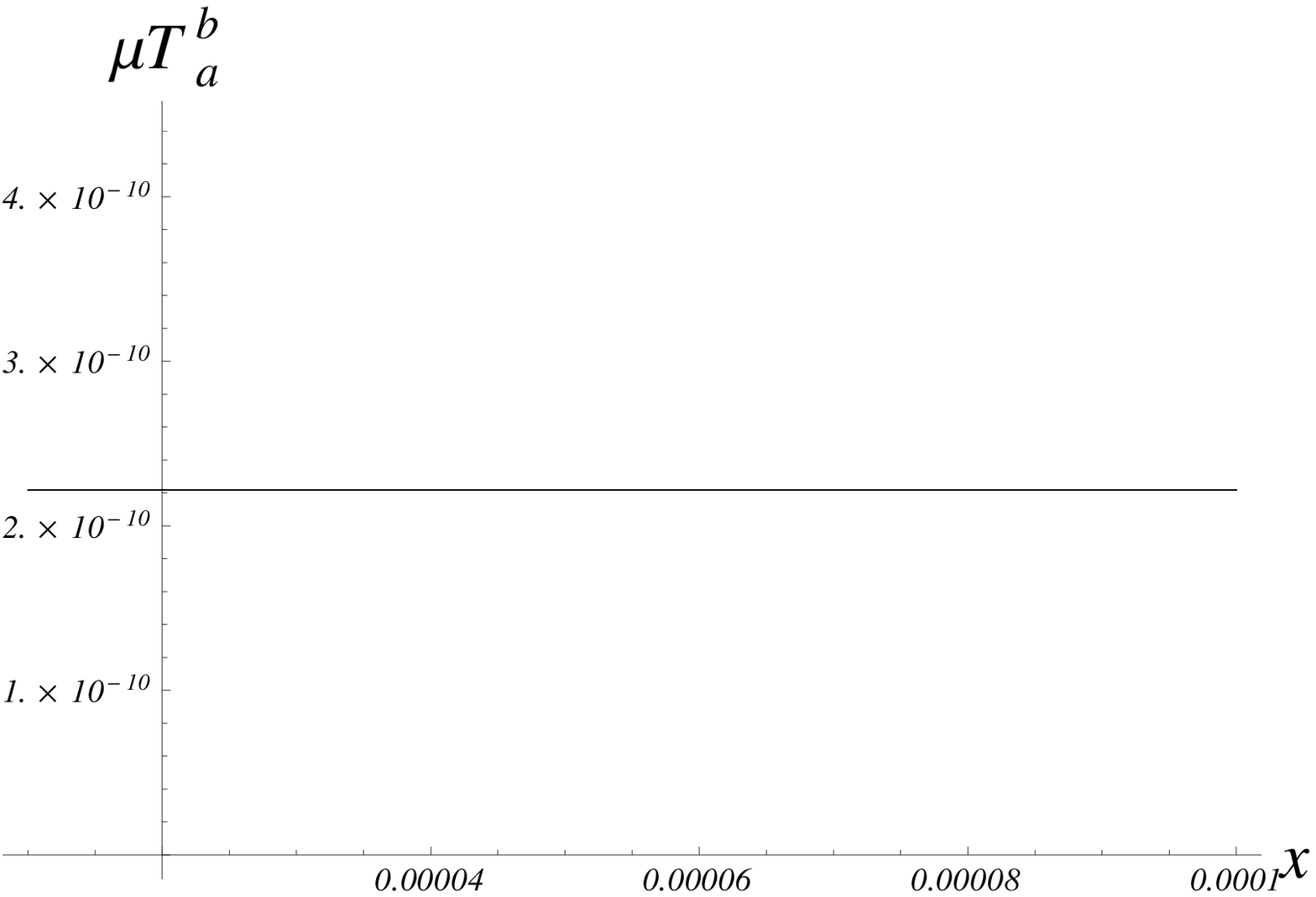}
\caption{The rescaled components of the stress-energy tensor of the massive conformally-coupled
scalar field in the closest vicinity of the regular center. The stress-energy tensor
coincides with the tensor in the de Sitter geometry. $[\mu =\mathcal{M}^{6}m^{2}/10^{2}]$. }
\label{pp2c}
\end{figure}

The general form of the stress-energy tensor is rather complicated,
and, for obvious reasons, it will not be presented here. 
(The components of the tensor $\bar{T}_{a}^{(i)b} $ for scalar, spinor and 
vector fields are available from the computer code repository~\cite{sklad}).
Instead, we shall analyze the components of the tensor numerically. 
First, let us consider the most interesting case of the ultraextremal and 
regular black hole. Such a configuration is characterized by $q=q_{crit}$ 
and $\ell=\ell_{crit}.$ The triply degenerated horizon is located 
at $x=x_{crit}.$  The run of the (rescaled) components of the tensor
of the quantized conformally coupled massive scalar field is displayed
in Figs.~\ref{pp2a}-\ref{pp2c}.

For the massive spinor and vector field the key features of the renormalized tensor 
are similar.
Inspection of the figures shows that although the components of the stress-energy tensor are 
small in the vicinity of the degenerate horizon and near the black hole center,
they are quite big in the region $0.03 < x < 0.25.$ Indeed, the ratio 
$|T_{a}^{b}|_{max}/|T_{a}^{b}|_{x_{crit}}\sim 10^{7}$
and  $|T_{a}^{b}|_{max}/|T_{a}^{b}|_{x=0}\sim 10^{9}$. 
Of course, this behavior is not unexpected and can be ascribed to the particular
form of the $f(r)$ function in that region. Simple consideration shows that the
leading terms are those with the lowest power of $\beta$ and highest of $x$ and
the oscillatory character of the components of the tensor is due to the
competition between various such terms in the region adjacent to the region 
described by the nearly flat metric potentials.
Fig.~\ref{pp2b} reveals existence of the layers of negative energy-density.
This behavior raises two questions.
First, is it permissible to use the semiclassical approximation inside the
degenerate horizon, and second, whether the perturbative approach is legitimate
in this region. The first question have been addressed to earlier, and the
affirmative answer can be given provided the Compton length associated with the
quantized field is much smaller that the characteristic radius of the curvature
of the background geometry. The second question is more subtle as the
perturbative approach usually involves asymptotic series and to answer it we
employed the general tensor calculated form the action functional constructed
from the  coincidence limit of the Hadamard-DeWitt coefficient $[a_{4}]$ as presented 
in Ref.~\cite{NextTo} and checked if these terms comprise small correction to 
the main approximation. We shall omit the details of the rather lengthy and not particularly 
illuminating calculations that we have carried out with the aid of the computer algebra. 
Once again the answer is affirmative, but now, it is necessary to carefully examine the 
leading terms of the approximation, and, for a given black hole mass, to determine 
the minimal allowable mass of the field.

\begin{figure}[h]
\includegraphics[width=11cm]{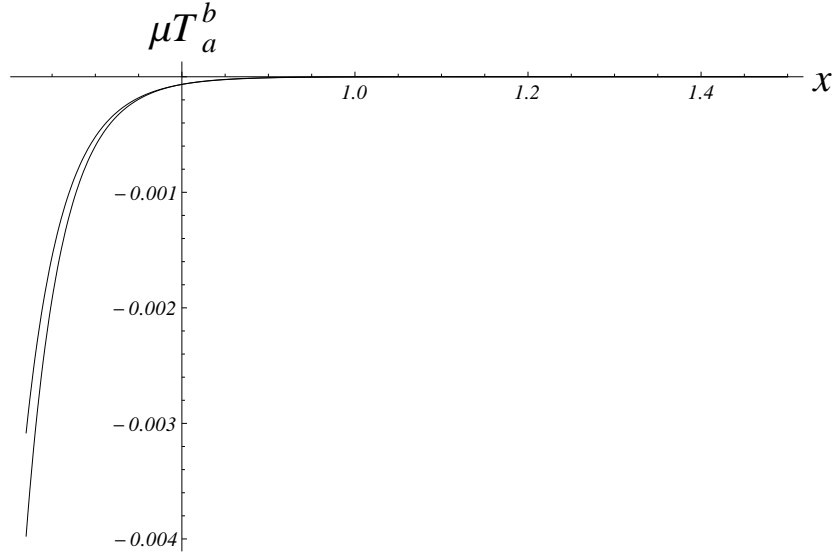}
\caption{The rescaled  components of the stress-energy tensor 
of the conformally coupled massive scalar field inside the ultraextremal 
Reissner-Nordstr\"om-deSitter black hole. Top to 
bottom at $x=0,1$  the curves represent the time and  angular  components
of the stress-energy tensor, respectively $[\mu =\mathcal{M}^{6}m^{2}/10^{2}]$. }
\label{rnds1}
\end{figure}

\begin{figure}[h]
\includegraphics[width=11cm]{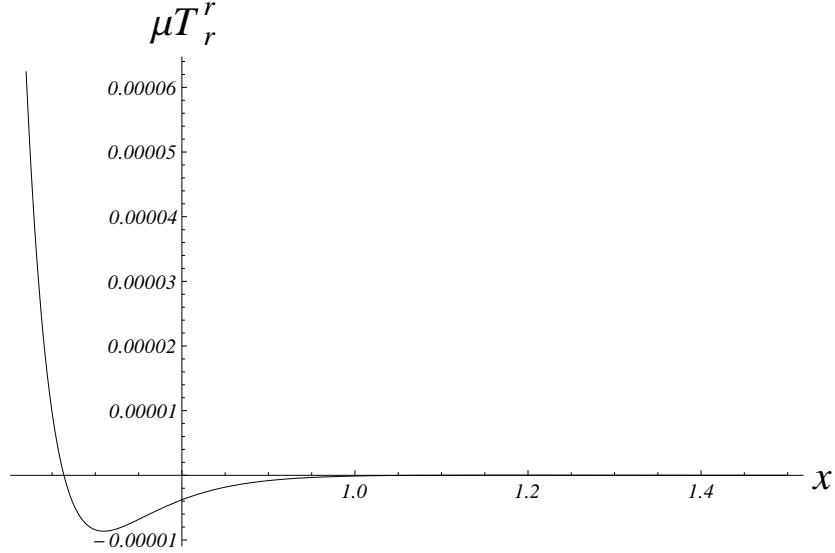}
\caption{The rescaled radial component of the stress-energy tensor of the massive 
conformally-coupled scalar field inside the ultraextremal Reissner-Nordstr\"om-de Sitter
black hole. $[\mu =\mathcal{M}^{6}m^{2}/10^{2}]$. }
\label{rnds2}
\end{figure}

Finally, let us compare the stress-energy tensor of the quantized massive fields
with the analogous tensor calculated in the ultraextremal Reissner-Nordstr\"om-de Sitter
black hole.  The ultraextremal configuration is characterized by 
\begin{equation}
\mathcal{M} = \frac{2}{3} \rp, \hspace{0.5cm} Q^{2} = \frac{1}{2}\rp^{2}, \hspace{0.5cm}
l^{2} =\frac{1}{2} \rp^{-2}.
\label{crit_RNDS}
\end{equation}
and the line element is given by (\ref{el_gen}) with $\psi(r) =0$ and
\begin{equation}
f(r) = -\frac{r^{2}}{6 \rp^{2}}\left( 1-\frac{\rp}{r}\right)^{3}\left(1+\frac{3\rp}{r} \right).
\end{equation}
After some algebra one has
\begin{equation}
\bar{T}_{a}^{(i) b} = \sum_{n=4}^{12} [\beta^{(i)}_{n}]_{a}^{b} \frac{\rp^{n-6}}{r^{n}} + 
\bar{ \mathfrak{T}}_{a}^{(i)b}, 
\end{equation}
with $[\beta^{i}_{5}]_{a}^{b} =0,$
where the coefficients $\beta$ depend on the spin of the field and (for the massive scalar
field)  the coupling constant $\xi.$ The tensor $\bar{\mathfrak{T}}_{a}^{(i)b}$ is given by 
(\ref{deS}) with the critical $l$ (See Eq.~\ref{crit_RNDS}). 
Since the stress-energy tensor diverges as $r \to 0$ it is expected that its applicability
is severely limited: It gives reasonable results in the region close to the  event horizon.
The components of the tensor are available from~\cite{sklad}.

\section{Conclusions and discussion}

In this paper we analyzed  a three-parameter class of static and 
spherically symmetric regular black holes
in the (anti-)de Sitter universe with the special emphasis
put on their horizon structure and the geometry inside the degenerate 
horizons. Each configuration is uniquely specified by the mass, the
(magnetic) charge and the cosmological constant. The cosmological constant 
is treated on the same footing as other fundamental constants.

We have studied the regularized stress-energy tensor of the quantized 
massive spinor, scalar and vector fields in a large mass limit within 
the framework of the Schwinger-DeWitt formalism inside the event horizon 
of the (ultra)extremal regular black hole. It is shown that the stress-energy
tensor exhibits oscillatory behaviour with the layers of the negative 
energy-density and the amplitudes in the region $0.03 < x < 0.25$
are a few orders of magnitude greater than its  components  near the 
event horizon or the  regular center. It can be demonstrated 
(making use of the computer codes \cite{sklad}) that this behavior persists  
for a wide range of ABGB(a-)dS parameters. This is an interesting and important 
behaviour, especially in the context of the back-reaction of the quantized 
fields upon the black hole geometry. On the other hand, the stress-energy
tensor in the vicinity of the event horizon, depending on which horizons 
do merge, coincides with  the tensor calculated in the Bertotti-Robinson, 
Nariai and Pleba\'nski-Hacyan spacetime. For completeness we have also 
calculated the stress-energy tensor for six physical product geometries.
Similarly, near the regular center the stress-energy tensor can be 
approximated by the tensor calculated in the (anti-)de Sitter spacetime. 

The ultraextremal ABGB-dS black hole provides a natural setting 
for analysis of the degenerate horizon under the influence of the 
quantized fields. In view of the results obtained in Refs.~\cite{Lowe,oleg_i_ja,Berej,jirinek01b}
it seems that it would be possible to construct the self-consistent
semiclassical analog of the classical ultraextremal ABGB-dS black 
hole.  Similarly, the important problem of the characteristics 
of the quantized fields in the Schwarzschild black hole in the asymptotically 
anti-de Sitter geometries has not been touched upon in this paper. It would 
be interesting to supplement and extend the discussion of Ref.~\cite{tanaka1,tanaka2}. 
We intend to return to this group of problems elsewhere.


\end{document}